\documentclass[epj,nopacs]{svjour}
\usepackage{axodraw}
\usepackage{epsfig}
\usepackage{amsfonts}
\usepackage{amsmath}
\usepackage{bbm}

\input pix.sty

\renewcommand{\eq}{eq.~}

\renewcommand{\fig}{fig.~}

\def\lsi{\raise0.3ex\hbox{$<$\kern-0.75em\raise-1.1ex\hbox{$\sim$}}}
\def\gsi{\raise0.3ex\hbox{$>$\kern-0.75em\raise-1.1ex\hbox{$\sim$}}}
\newcommand{\lsim}{\mathop{\lsi}}
\newcommand{\gsim}{\mathop{\gsi}}

\newcommand{\rmii}[1]{{\mbox{\tiny\rm{#1}}}}
\newcommand{\re}{\mathop{\mbox{Re}}}
\newcommand{\im}{\mathop{\mbox{Im}}}

\newcommand{\Tint}[1]{{\hbox{$\sum$}\!\!\!\!\!\!\!\int\,}_{\!\!\!\!\raise-0.9ex\hbox{$\scriptstyle{#1}$}}}
\newcommand{\Tinti}[1]{{{\Sigma}\!\!\!\!\raise0.3ex\hbox{$\int$}_\rmii{${#1}$}}}

\newcommand{\bi}{\begin{itemize}}
\newcommand{\ei}{\end{itemize}}

\renewcommand{\tt}{\tilde t_0}

\newcommand{\hide}[1]{ }

\def\TAsc(#1,#2)(#3,#4,#5)%
{\SetWidth{2.0}\CArc(#1,#2)(#3,#4,#5)\SetWidth{1.0}}
\def\Lwidth{3}

\def\TAgl(#1,#2)(#3,#4,#5){\SetWidth{2.0}\PhotonArc(#1,#2)(#3,#4,#5){\Lwidth}%
{6.283 #3 mul 360 div #4 #5 sub #4 #5 sub mul sqrt mul Tdensity mul}%
\SetWidth{1.0}}
\def\TLgl(#1,#2)(#3,#4){\SetWidth{2.0}\Photon(#1,#2)(#3,#4){\Lwidth}
{#1 #3 sub #1 #3 sub mul #2 #4 sub #2 #4 sub mul add sqrt Tdensity mul}%
\SetWidth{1.0}}
\newcommand{\piC}[1]{\;\parbox[c]{40pt}{\begin{picture}(120,60)(0,-20)
\SetWidth{1.0}\SetScale{0.35} #1 \end{picture}}\;}
\def\ConnectedA(#1,#2,#3){\piC{#1(60,-15)(75,34,146) #2(60,75)(75,214,326)%
 #3(60,60)(20,190,350)%
 \GBoxc(0,30)(10,10){1} \GBoxc(120,30)(10,10){1}%
  }}
\def\ConnectedB(#1,#2,#3){\piC{#1(60,-15)(75,34,146) #2(60,75)(75,214,326)%
 #3(60,60)(60,0)%
 \GBoxc(0,30)(10,10){1} \GBoxc(120,30)(10,10){1}%
  }}
\def\ConnectedC(#1,#2){\piC{#1(60,-15)(75,34,146) #2(60,75)(75,214,326)%
 \GBoxc(0,30)(10,10){1} \GBoxc(120,30)(10,10){1}%
  }}
\def\ConnectedD(#1,#2){\piC{#1(60,-15)(75,34,146) #2(60,75)(75,214,326)%
 \GBoxc(0,30)(10,10){1} \GBoxc(120,30)(10,10){1}%
 \SetWidth{2.0} 
 \Line(55,55)(65,65)%
 \Line(55,65)(65,55)
  }}

\begin{document}
\sloppy


\title{A test on analytic continuation of thermal imaginary-time data}

\author{
Y.~Burnier\inst{1}
  \and 
M.~Laine\inst{2}
  \and 
L.~Mether\inst{2}
}

\institute{
Dept.\ of Physics and Astronomy, 
SUNY, Stony Brook, New York 11794, USA
 \and
Faculty of Physics, University of Bielefeld, 
D-33501 Bielefeld, Germany}

\date{Received: 3 February 2011 / Revised: 8 March 2011}

\abstract{%
Some time ago, Cuniberti {\it et al} have proposed a novel 
method for analytically continuing thermal imaginary-time 
correlators to real time, which requires no model input and 
should be applicable with finite-precision data as well. Given 
that these assertions go against common wisdom, we report on 
a naive test of the method with an idealized example. We do 
encounter two problems, which we spell out in detail; 
this implies that systematic errors are difficult to 
quantify. On a more positive note, the method 
is simple to implement and allows for an empirical
recipe by which a reasonable qualitative estimate for some 
transport coefficient may be obtained, if statistical errors of 
an ultraviolet-subtracted imaginary-time measurement 
can be reduced to roughly below the per mille level. 
}


\maketitle

%
\section{Introduction}

%
It is a perennial problem that a reliable study of strong interactions
at temperatures around a few hundred MeV requires non-perturbative
methods, yet standard Monte Carlo techniques only work in Euclidean 
signature. This implies that many of the physically most interesting 
observables, such as transport coefficients or particle production 
rates (for reviews see e.g.\ refs.~\cite{ga_rev,hbm_rev}), which 
are inherently Minkowskian in nature, are difficult to address. 

A few years ago, a possible solution 
to this problem was proposed~\cite{hatsuda}, 
through the use of a ``Maximum Entropy Method''~\cite{mem}.
Unfortunately, despite many implementations and various  
related attempts (see e.g.\ refs.~\cite{sg}--\cite{rbc}; similar
discussions continue also in the context of condensed matter 
physics problems, 
see e.g.\ ref.~\cite{condmat2} and references therein), 
it remains a problem that there appears to be uncontrolled 
dependence on model input in these results. Therefore, 
it is perhaps worthwhile to look for alternatives as well.

Some time ago, Cuniberti {\em et al}~\cite{cuniberti} put forward
a concrete suggestion which could help in this respect. An algorithm
was provided which was shown to yield a correct analytic
continuation in a specific limit; furthermore, it was suggested that
the method might work even if there is a finite amount of data and 
the data points have error bars. Despite these attractive features, 
we are not aware of a previous numerical implementation of the 
algorithm. For the record,
we wish to report one in this short note, even if quantitative 
success is somewhat marginal. Our hope is that some of our theoretical 
remarks are nevertheless of interest and that, on the qualitative level, 
the algorithm might turn out to be 
a useful addition to the existing tool kit. 


%
\section{Basic idea}

%
The general philosophy of the method
can be summarized as follows. We consider a periodic 
(``bosonic'') Euclidean correlator, ${\mathcal G}(\tau,\cdot)$, 
with ${\mathcal G}(\tau+k\beta,\cdot) = {\mathcal G}(\tau,\cdot)$
where $k\in \mathbbm{Z}$, $\beta$ denotes the inverse temperature, 
and $\cdot$ denotes suppressed variables such as spatial momentum. 
The correlator should be analytic everywhere except at 
$\re\tau = 0 + k\beta$; there it should still be continuous. 
We also assume the further property  
${\mathcal G}(\beta-\tau,\cdot) = {\mathcal G}(\tau,\cdot)$, 
$0 < \tau < \beta$,   
which was not imposed in ref.~\cite{cuniberti} but is satisfied
by typical gauge-invariant current-current 
correlators measured in lattice QCD. 
Given the finiteness of $\beta$, the
Fourier representation involves a discrete
set of Matsubara frequencies, 
$
 \tilde {\mathcal G}(\omega_n,\cdot) \equiv
 \int_0^\beta \! {\rm d}\tau \, e^{i \omega_n\tau } 
 {\mathcal G}(\tau,\cdot)
$, 
$\omega_n = 2 \pi n T$, 
$n \in \mathbbm{Z}$, $T \equiv \beta^{-1}$. 
Making use of a general hypothesis about the asymptotic behaviour
of various correlators in the Minkowskian time domain, namely that they  
show at most powerlike growth at infinity
(physically relevant correlators are actually expected
to vanish at infinity, see e.g.\ ref.~\cite{ay}), 
a unique analytic continuation to a part of the complex plane 
can be shown to exist and can be constructed explicitly. 

%
\begin{figure*}

\hspace*{1.0cm}%
\begin{minipage}[c]{6.5cm}
\begin{picture}(160,160)(-80,-80)
\SetScale{1.0}  
\SetWidth{0.5}
\Line(7,0)(7,75)%
\Line(-7,0)(-7,75)%
\CArc(0,0)(7,180,360)%
\ArrowLine(7,65)(7,75)%
\ArrowLine(-7,75)(-7,65)%
\BBox(-97,67)(-63,83)%
\SetWidth{1.0}
\LongArrow(-80,0)(80,0)%
\LongArrow(0,-80)(0,80)%
\ZigZag(0,0)(0,70){2}{16}%
\ZigZag(-50,0)(-50,70){2}{16}%
\ZigZag(50,0)(50,70){2}{16}%
\Text(75,75)[c]{$\tau$}
\Line(70,70)(80,70)%
\Line(70,70)(70,80)%
\Text(-80,75)[c]{$\cal{G}^+(\tau)$}%
\Text(12,70)[l]{$t$}%
\Text(11,55)[l]{$J^+(t)$}%
\Text(-55,-10)[c]{$-\beta$}%
\Text(50,-10)[c]{$\beta$}%
\end{picture}
\end{minipage}%
\hspace*{1.0cm}%
\begin{minipage}[c]{6.5cm}
\begin{picture}(160,160)(-80,-80)
\SetScale{1.0}  
\SetWidth{0.5}
\CBox(-80,-80)(0,80){Cyan}{Cyan}
\Line(7,75)(7,-75)%
\ArrowLine(7,-65)(7,-75)%
\Line(32,-3)(38,3)%
\Line(32,3)(38,-3)%
\Line(67,-3)(73,3)%
\Line(67,3)(73,-3)%
\BBox(-97,67)(-63,83)%
\SetWidth{1.0}
\LongArrow(-80,0)(80,0)%
\LongArrow(0,-80)(0,80)%
\Text(75,76.5)[c]{$\zeta$}
\Line(70,70)(80,70)%
\Line(70,70)(70,80)%
\Text(-80,75)[c]{$\tilde{J}^+(\zeta)$}%
\Text(12,-70)[l]{$\omega$}%
\Text(11,-55)[l]{$\tilde{R}(\omega)$}%
\Text(35,-10)[c]{$2\pi T$}%
\Text(70,-10)[c]{$4\pi T$}%
\end{picture}
\end{minipage}%

\caption{
The complex planes, basic functions, and 
analytic structures as discussed in the text. 
} 
\la{fig:planes}
\end{figure*}
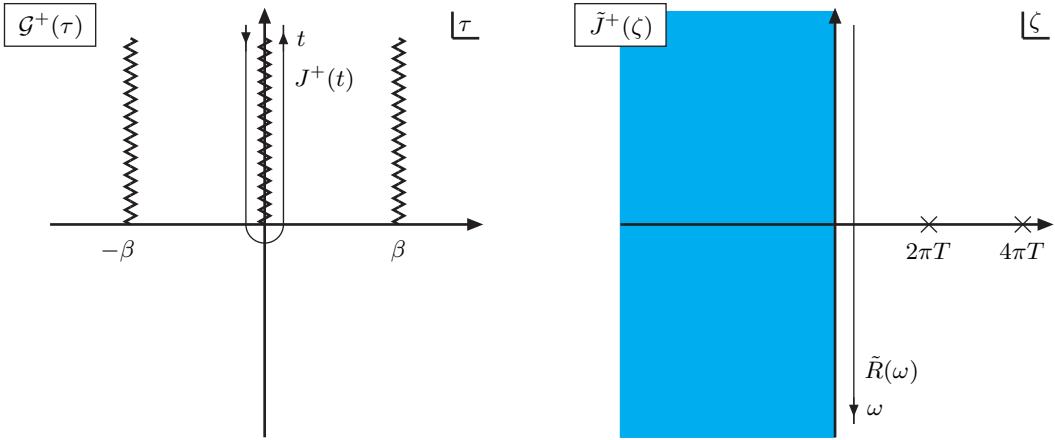
%

More concretely,\footnote{%
 The following relations, though simple to state, 
 may not be entirely obvious at first sight; 
 mathematical proofs 
 can be found in ref.~\cite{cuniberti}
 and references therein. 
 } 
the basic quantities defined 
in ref.~\cite{cuniberti} are the one-sided sum 
$
 {\mathcal G}^+(\tau,\cdot) \equiv 
 T \sum_{\omega_n \ge 0} \tilde {\mathcal G}
 (\omega_n,\cdot) e^{-i \omega_n \tau} 
$, 
which is analytic for 
$\im\tau < 0$ but has cuts in the upper half-plane; 
its discontinuity across the cut starting at the origin, 
$
 J^+(t,\cdot) \equiv i \bigl[ {\mathcal G}^+(\epsilon + i t,\cdot) - 
 {\mathcal G}^+(-\epsilon + i t,\cdot) \bigr]
$, $t>0$, $\epsilon = 0^+$, 
which equals the retarded real-time correlator, 
${\mathcal R}(t,\cdot)$; 
as well as its Laplace transform 
$
 \tilde J^+(\zeta,\cdot) \equiv
 \int_0^\infty \! {\rm d}t \, e^{-\zeta\, t} J^+(t,\cdot)
$, 
which is analytic for $\re \zeta > 0$. 
For $\zeta = \omega_n$, $\tilde J^+(\zeta,\cdot)$
reduces to the Fourier components $\tilde {\mathcal G}(\omega_n,\cdot)$,
and therefore constitutes the desired analytic continuation 
to a complex half-plane. 
The value of $\tilde J^+(\zeta,\cdot)$
along the axis $\zeta = \epsilon -i \omega $, 
$\omega \in \mathbbm{R}$, 
yields the Fourier transform of the retarded correlator, 
$\tilde{\mathcal R}(\omega,\cdot)$, 
whose imaginary part in turn equals the spectral function. 
The basic analytic structure is illustrated in  \fig\ref{fig:planes}.


%
\section{Algorithm}
\la{Algorithm_idealized}

%
To implement the analytic continuation, the idea of ref.~\cite{cuniberti}
is to expand $\tilde{J}^+(\zeta,\cdot)$ with the help of 
Pollaczek polynomials (of the type defined on an infinite interval);
the retarded correlator $J^+(t,\cdot)$ is in turn expressed as a linear 
combination of Laguerre polynomials, with argument $2 e^{-2\pi t T}$. 
More explicitly, taking ref.~\cite{cuniberti} at face value, 
the steps are as follows: 

\begin{itemize}

\item[(i)] 
Compute the Fourier modes, 
$\tilde {\mathcal G}(\omega_n,\cdot)$. Due to 
${\mathcal G}(\beta-\tau,\cdot) = {\mathcal G}(\tau,\cdot) 
\in \mathbbm{R}$, the Fourier modes are real 
and even in $\omega_n\to -\omega_n$. 

\item[(ii)] 
Construct the coefficients
\ba
 a_\ell & \equiv & 
 2 (-1)^\ell \sum_{n=0}^\infty
 \frac{(-1)^n}{n!} \tilde {\mathcal G}(\omega_{n+1},\cdot) \, 
 _2 F_1(-\ell,n+1;1;2)
 \;, \nn 
 & & \hspace*{4cm} \ell = 0, 1, 2, \ldots 
 \;. \la{al_ideal}
\ea
We have re-expressed the Pollaczek polynomials of 
ref.~\cite{cuniberti}
through the hypergeometric function $_2 F_1$.\footnote{%
 More precisely: 
 $
  P_\ell(\eta) \equiv P_\ell^{1/2}(\eta;{\frac{\pi}{2}})
  = i^\ell \, _2 F_1 (-\ell,\fr12 + i \eta; 1; 2)
 $, cf.\ ref.~\cite{book}.
 }
Note that the Matsubara zero-mode does not contribute in \eq\nr{al_ideal}.

\item[(iii)] 
According to ref.~\cite{cuniberti}, 
$a_\ell$ decreases with $\ell$, such that 
 $\sum_{\ell=0}^\infty |a_\ell|^2$
is finite. 

\item[(iv)] 
Defining $\tilde t \equiv 2 \pi t T$, 
the retarded real-time correlator can now be obtained as 
\be
 J^+(\tilde t,\cdot) = e^{-e^{-\tilde t}}
 \sum_{\ell = 0}^{\infty} a_\ell \, L_\ell(2 e^{-\tilde t})
 \;, \la{Jplus_ideal}
\ee
where $L_\ell$ are the Laguerre polynomials. 
Given that $L_\ell(0) = 1$, the asymptotic value
is $J^+(\infty,\cdot) = \sum_{\ell=0}^\infty a_\ell$, 
which should vanish in physically meaningful examples~\cite{ay}. 

\item[(v)]
Given $J^+$, the spectral correlator reads 
$
 \rho(t) = \frac{1}{2i} \bigl[ \theta(t) J^+(t) - 
 \theta(-t) \bar{J}^+(-t) \bigr]
$, 
where $\bar{J}^+$ denotes a complex conjugate. 
Noting that $J^+$ as produced by \eq\nr{Jplus_ideal}
is real under our assumptions, 
the spectral function can finally be obtained as
\be
 \rho(\tilde \omega,\cdot)
 = \int_0^\infty 
 \! {\rm d}\tilde t \, 
 \sin (\tilde\omega \tilde t)\, J^+(\tilde t,\cdot)
 \;,
 \quad
 \tilde \omega \equiv \frac{\omega}{2\pi T}
 \;. 
\ee

\end{itemize}


%
\section{Practical implementation}
\la{Algorithm_in_practice}

%
To apply the previous algorithm to a situation where 
only a finite number of data points are available, we adopt 
the following steps, with the same numbering as in section~3. 

\begin{itemize}

\item[(i)] 
We assume the interval $0 < \tau < \beta$ to be evenly
divided into $N$ parts. The Fourier modes are constructed by
a discrete transformation; we leave out the ``contact point''
($\tau = 0$ or $\tau = \beta$), given its possible 
inaccessibility to 
practical measurement. Then, 
\ba
 \tilde {\mathcal G}(\omega_n,\cdot) 
 & \simeq & 
 {\textstyle \frac{\beta}{N} } 
 \sum_{k=1}^{N-1}
 e^{i \frac{2\pi n k}{ N}}
 {\mathcal G}\Bigl(
 {\textstyle \frac{k\beta}{N} } 
 ,\cdot\Bigr)
 \;, \nn
 & & \hspace*{2cm}
 n = 0,\ldots,N-1
 \;. \la{fourier_real}
\ea
We stress again that, thanks to 
${\mathcal G}(\beta-\tau,\cdot) = {\mathcal G}(\tau,\cdot) \in \mathbbm{R}$ 
for $0 < \tau < \beta$, 
$\tilde {\mathcal G}(\omega_n,\cdot) \in \mathbbm{R}$.

\item[(ii)] 
Defining the coefficients
$
 \phi_{\ell,n} \equiv \frac{(-1)^\ell}{n!} \, 
  _2 F_1(-\ell,n+1;1;2)
$, 
$n\ge 0$, 
which can be constructed from the recurrence relation
\ba
 && \phi_{\ell,-1} \equiv 0
 \;; \quad
 \phi_{\ell,0}  = 1
 \;, \nn 
 && 
 \phi_{\ell,n} = \frac{(2\ell + 1)\phi_{\ell,n-1} + \phi_{\ell,n-2}}{n^2}
 \;, \quad
 n \ge 1
 \;, \la{recurrence}
\ea
the $a_\ell$ can be obtained from 
\be
 a_\ell \simeq
 2 \sum_{n=0}^{N-2}
 {(-1)^n} \tilde {\mathcal G}(\omega_{n+1},\cdot) \, \phi_{\ell,n}
 \;. \la{al_real}
\ee
The precise upper limit 
(be it $N-2$ or e.g.\ $N/2$) has little importance, 
because $\phi_{\ell,n}$ turn out to decrease rapidly
for $n > n_\rmi{max}$, where in general
$n_\rmi{max} \ll N/2$ (cf.\ below). 

\item[(iii)] 
According to ref.~\cite{cuniberti}, 
the $|a_\ell|$ decrease only up to some $\ell_\rmi{max}$, meaning that 
 $\sum_{\ell=0}^{\ell_\rmii{max}} |a_\ell|^2$
shows a plateau as a function of $\ell_\rmi{max}$, 
but eventually it diverges. Ref.~\cite{cuniberti} suggests
choosing $\ell_\rmi{max}$ from this plateau.
We find, however, 
that in practice it is more useful to monitor 
the sum $\sum_{\ell=0}^{\ell_\rmii{max}} a_\ell$; 
the reason is discussed in connection with \eq\nr{sumal1} below. 

\item[(iv)] 
Choosing $\ell_\rmi{max}$ according to some criterion, 
$J^+$ can be approximated through  
\be
 J^+(\tilde t,\cdot) \simeq e^{-e^{-\tilde t}}
 \sum_{\ell = 0}^{\ell_\rmii{max}} a_\ell \, L_\ell(2 e^{-\tilde t})
 \;, \la{Jplus_real}
\ee
where, as usual, the $L_\ell$ can be constructed from 
\ba
 && L_0(x) = 1\;, \quad
    L_1(x) = 1 - x  \;, \nn  
 &&
 L_\ell(x) = 2 L_{\ell -1}(x) - L_{\ell - 2}(x)
 \nn & & \hspace*{1cm} - 
 \frac{(1+x) L_{\ell -1}(x) - L_{\ell - 2}(x)}{\ell}
 \;, \quad \ell \ge 2
 \;. 
\ea 
In general the asymptotic value, 
\be 
 J^+(\infty,\cdot) = \sum_{\ell=0}^{\ell_\rmii{max}} a_\ell
 \;, \la{sumal1}
\ee 
does not vanish. In contrast, physically relevant  
current-current correlators should vanish at infinite time 
separation~\cite{ay}. It turns out that this supplementary 
information can be used to choose  
values of $\ell_\rmi{max}$, namely those for 
which $J^+(\infty,\cdot)$ vanishes 
approximately, offering ``windows of opportunity'', in which the 
algorithm appears to perform reasonably well even with non-ideal data. 

\item[(v)]
The spectral function can be obtained as
\be
 \rho(\tilde \omega,\cdot)
 \simeq \int_0^\infty 
 \! {\rm d}\tilde t \, 
 \sin (\tilde\omega \tilde t)\, 
 \Bigl[ J^+(\tilde t,\cdot) - J^+(\infty,\cdot) \Bigr] 
 \;, 
\ee
where we have subtracted by hand any
possible ``remnant'' $J^+(\infty,\cdot)$.
(Otherwise $ \rho(\tilde \omega,\cdot)$ would diverge as $\sim 1/\tilde\omega$
 at small frequencies.)

\end{itemize}
As the proofs in ref.~\cite{cuniberti} show, in the limit 
$N\to \infty$ and vanishing errors these steps do yield 
the correct spectral function for current-current correlators
of the considered type. 


%
\section{Problems}
\la{Problems}

%
We now turn to two problems that limit 
the usefulness of the algorithm specified above. For simplicity
the discussion will be carried out in the combined continuum 
and infinite-volume limit; a finite cutoff alleviates problem (i) 
but adds more structure to the spectral function and thereby 
renders problem (ii) worse. A finite volume can in principle 
also change the behaviour of Euclidean correlators in a physically
interesting way (cf.\ ref.~\cite{ll} and references therein), 
however we wish to ignore these effects for now. 
(Formally a smooth infinite-volume type shape could be obtained e.g.\  
by considering a suitable Gaussian smoothing of $\rho(\omega,\cdot)/\omega$.)


\begin{itemize}

\item[(i)] 
In order for the recipe to apply, we must 
be able to compute the Fourier coefficients 
$\tilde {\mathcal G}(\omega_n,\cdot)$; 
this implies that ${\mathcal G}(\tau,\cdot)$ must be integrable 
around $\tau = 0$ mod $\beta$. In fact, as mentioned above, 
in ref.~\cite{cuniberti} 
${\mathcal G}(\tau,\cdot)$ was even assumed to be 
continuous (and therefore finite) at $\tau = 0$ mod $\beta$.
In contrast, the correlation functions of composite operators
that are relevant for the determination of transport coefficients
or particle production rates in QCD diverge 
at small $\tau$ in the continuum limit. 

In principle, a possible way around the problem might 
be to consider the temperature derivative  of 
a correlator, rather than a correlator as such. This could
work if the derivative is taken in fixed physical units. 
Sometimes, such differences are rather taken in scaled
units, i.e.\ by subtracting values of 
$\hat{\mathcal G}(\hat\tau,\cdot) \equiv
\beta^p {\mathcal G}(\hat\tau \beta,\cdot)$, 
$0 < \hat\tau < 1$, at different $\beta$'s, but such differences are 
continuous only if there are no scaling violations at small~$\tau$. 

In terms of a spectral function, the finiteness 
of ${\mathcal G}(0^+,\cdot)$ necessitates 
$\rho(\omega,\cdot) \!\le C / (\omega \ln^2\!\omega )$ 
at $\omega \gg T$, cf.\ \eq\nr{master}. The asymptotics
of various spectral functions in this regime have 
been analyzed in ref.~\cite{sch2}, and for vector current 
correlators the decay of the thermal part is indeed 
fast enough to satisfy the bound. 


\item[(ii)] 
As can be seen from \eq\nr{al_real}, the factor $(-1)^n$
implies a cancellation between Fourier modes
in the construction of the $a_\ell$'s; when the $a_\ell$'s
multiply the Laguerre polynomials in \eq\nr{Jplus_real}, 
further cancellations take place, particularly at large $\tilde t$
(cf.\ \eq\nr{sumal1}).
This implies a substantial significance loss. Furthermore, 
as can be seen from \eq\nr{recurrence}, for a fixed $\ell$ 
the coefficients $\phi_{\ell,n}$ grow very fast at small $n$, 
reaching a maximal value at $n_\rmi{max} \approx \sqrt{2(\ell +1)}$. 
The maximal value, $\phi_{\ell,n_\rmii{max}}$, grows with $\ell$, 
exceeding
$10^4$ at $\ell = 21$ 
(for $n_\rmi{max} = 6$), 
$10^6$ at $\ell = 40$ 
(for $n_\rmi{max} = 9$) and 
$10^8$ at $\ell = 64$
(for $n_\rmi{max} = 11$).
So, the Fourier coefficients around $n\sim n_\rmi{max}$
would need to be determined 
with the corresponding relative accuracy in order to obtain
a meaningful signal even after the cancellations.
This is obviously a formidable challenge, particularly 
considering that problem (i) already
requires a subtraction and inflicts
an associated significance loss. 

\end{itemize}


%
\section{Test}

\begin{figure*}


\centerline{%
\epsfysize=7.5cm\epsfbox{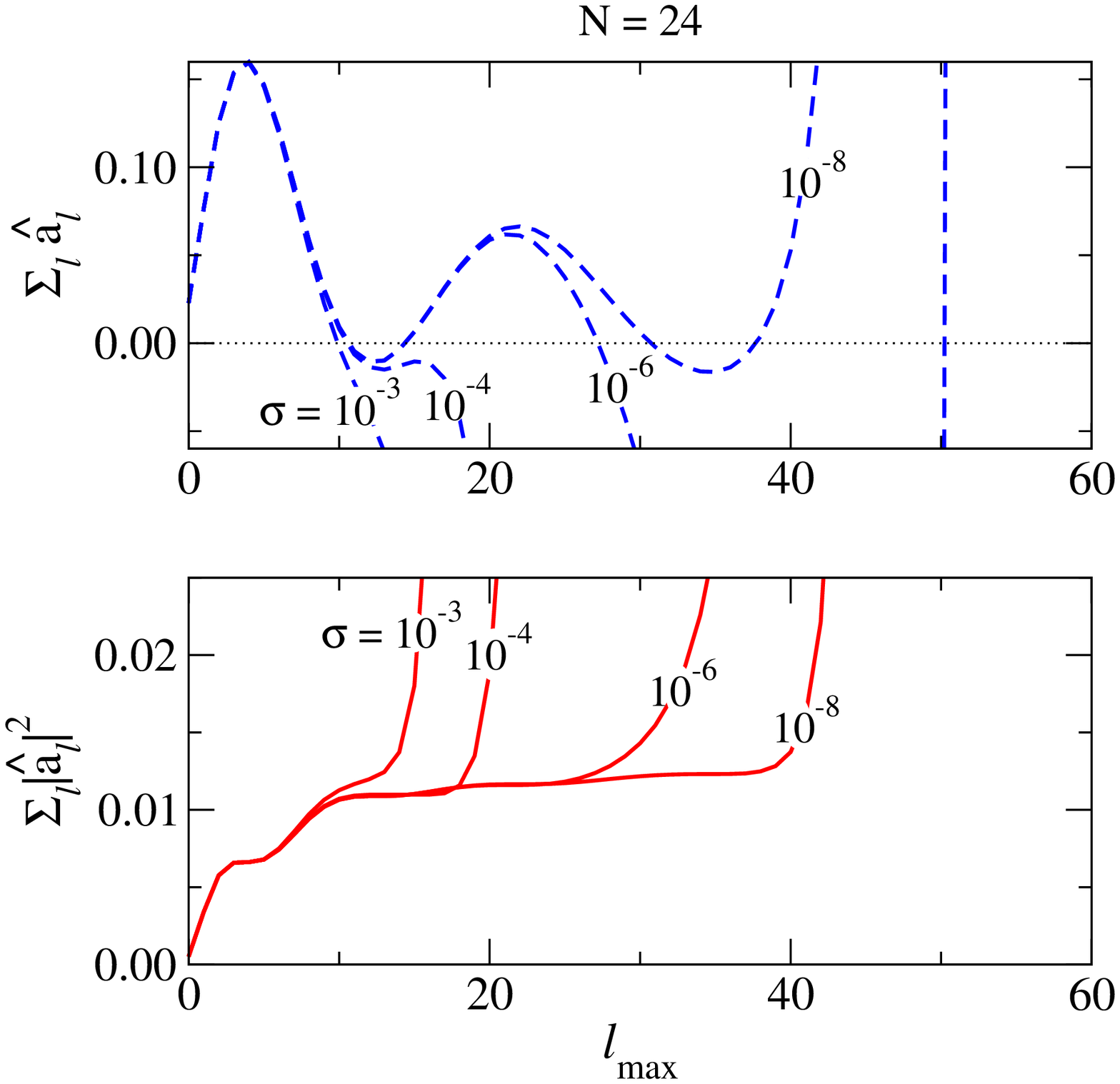}%
~~\epsfysize=7.5cm\epsfbox{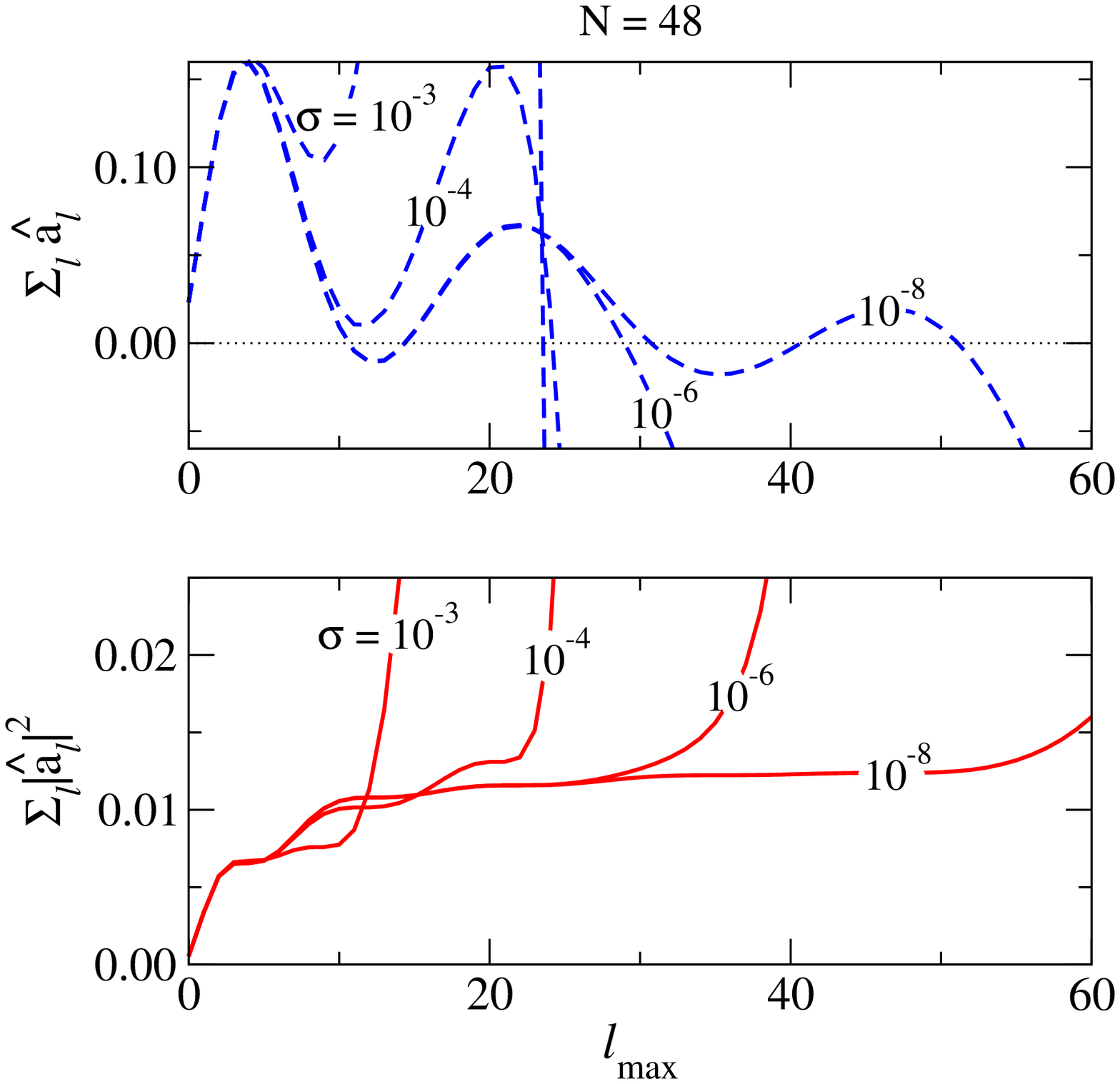}%
}

\caption{ 
 Top: $\sum_{\ell = 0}^{\ell_\rmii{max}} \hat a_\ell$ 
 as a function of $\ell_\rmi{max}$,
 for $N=24$ (left) and $N=48$ (right); here 
 $\hat a_\ell \equiv a_\ell / T^3$ and $\sigma$ 
 indicates the local relative standard deviation
 of ${\mathcal G}(\tau,\cdot)$
 (results are shown for one random configuration).  
 Bottom: the corresponding
 $\sum_{\ell = 0}^{\ell_\rmii{max}} |\hat a_\ell|^2$.   
 }
\la{fig:1}
\end{figure*}

%
Despite the problems of the previous section, we now wish 
to demonstrate that in principle the method may still work
on the qualitative level. In order
to achieve this, we assume that a suitable ultraviolet subtraction
has been carried out, and that our data are quite precise. 

As a model, we take inspiration from a spectral function related 
to an ``electric field'' correlator yielding the momentum-diffusion
coefficient of a heavy quark~\cite{ct,eucl,hbm2}. We assume 
a substantial positive intercept of $\rho(\omega,\cdot)/\omega$ 
at zero frequency (cf.\ ref.~\cite{chm}) and decreasing behaviour 
at large frequency, just fast enough to yield a continuous 
${\mathcal G}(\tau,\cdot)$: 
\be
 \rho_\rmi{test} \equiv \frac{C\, \tilde\omega}{(2+\tilde\omega^2)^2}
 \biggl[ 
   1 - \frac{\tilde\omega^2}{\ln^2(2+\tilde\omega^2)}
 \biggr]
 \;, \quad
 \tilde{\omega} \equiv \frac{\omega}{2\pi T}
 \;. \la{model}
\ee
This spectral function is not positive-definite in order to reflect 
the fact that a suitable ultraviolet subtraction has been carried out, 
and because a negative
$
 \rho \sim -{T^4}/ {(\omega\ln^2 \omega)}
$ 
is precisely the qualitative asymptotic behaviour 
found in perturbation theory~\cite{rhoE} 
(the logarithm squared assumes that the gauge coupling
is let to run with $\omega$).\footnote{%
 Note that the Lorentzian form 
 $\rho \sim \tilde{\omega}/(\tilde{\eta}^2+\tilde{\omega}^2)$
 would not decrease fast enough at large frequencies. 
 } 
The coefficient $C$
appears linearly in all steps so that, without
loss of generality, we set $C \equiv 4 T^3$ in the following, 
thereby normalizing $\rho_\rmi{test}/(\tilde{\omega} T^3 )$
to unity at $\tilde\omega \to 0$.

\begin{figure*}


\centerline{%
\epsfysize=7.5cm\epsfbox{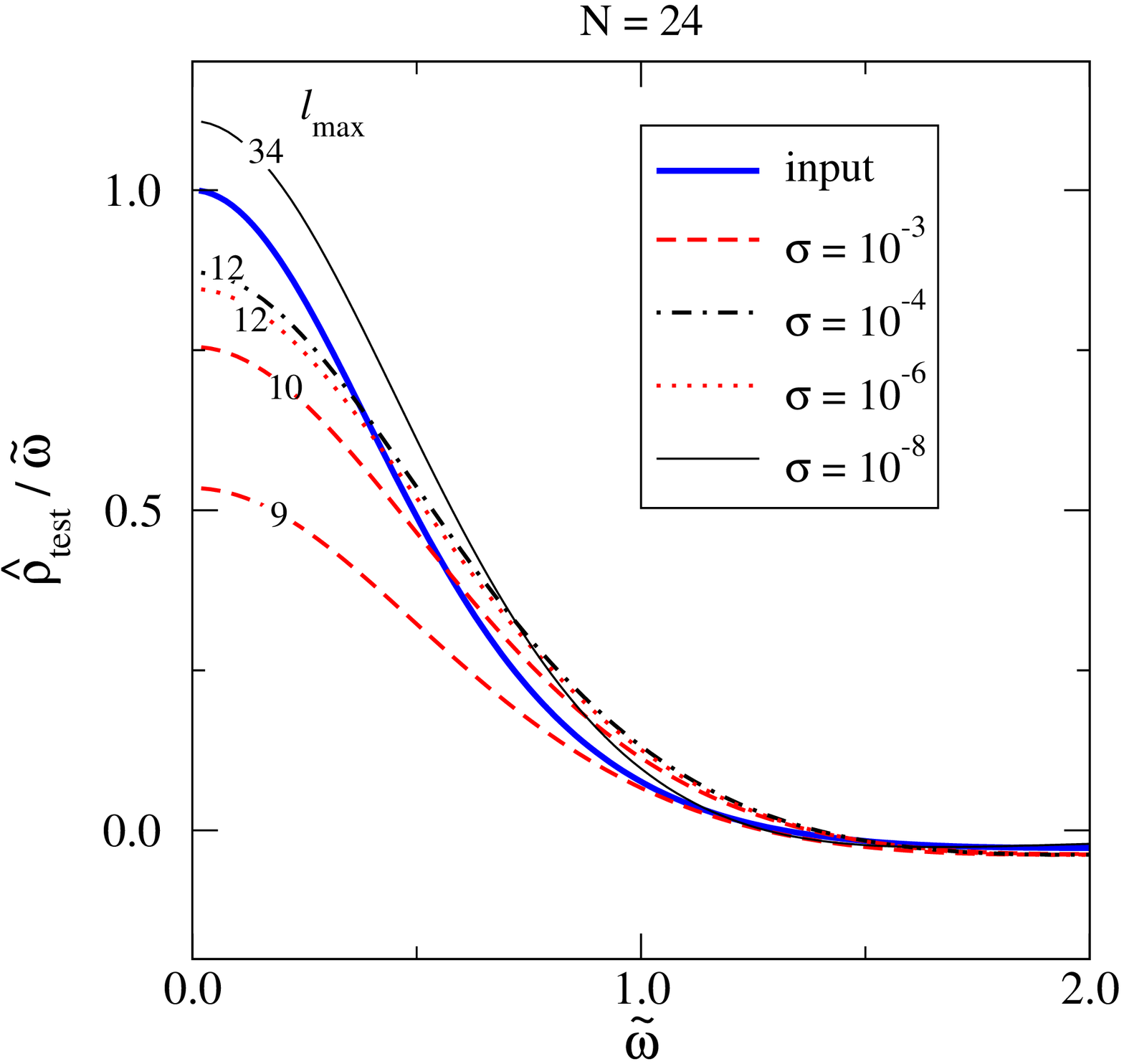}%
~~\epsfysize=7.5cm\epsfbox{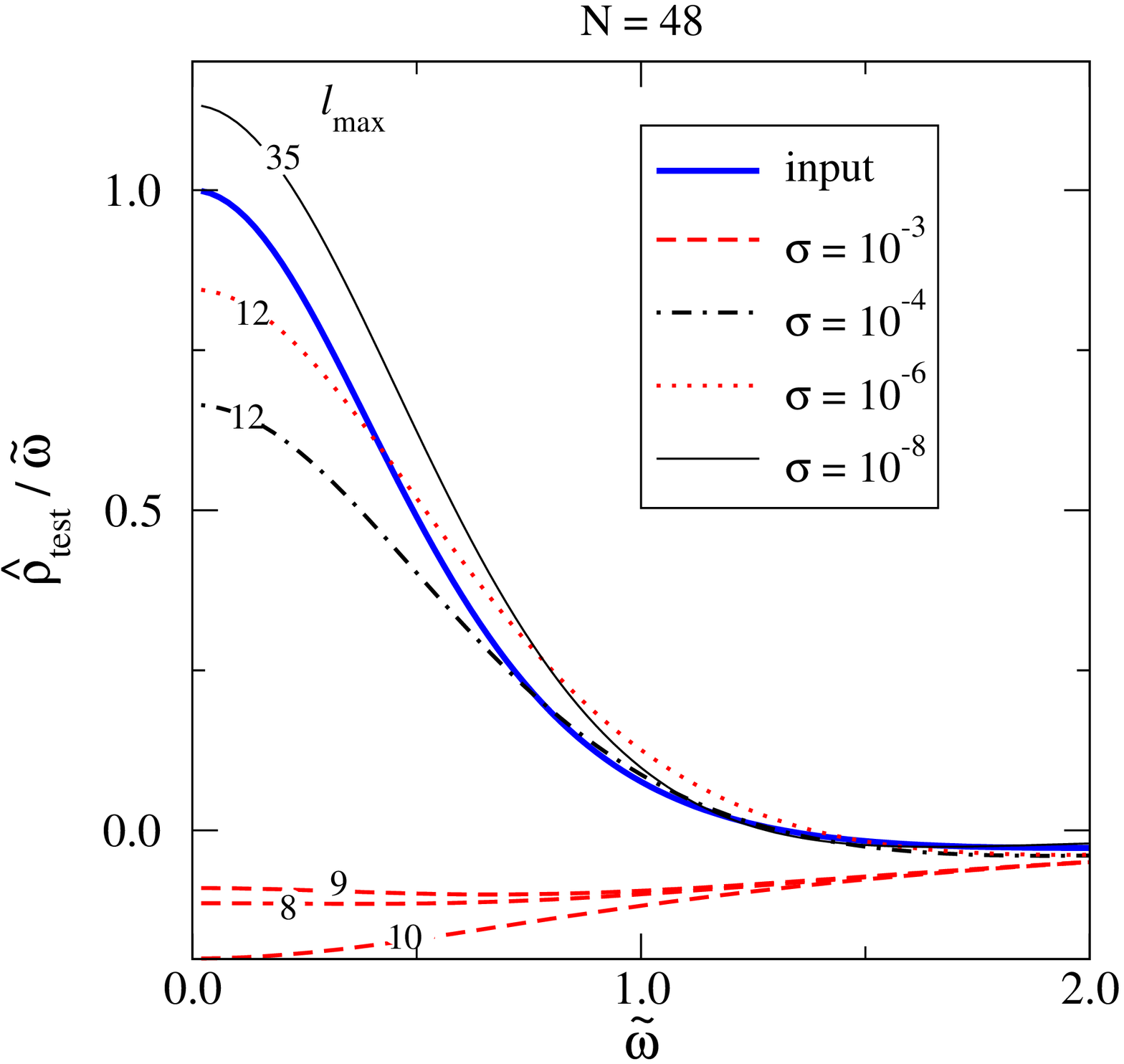}%
}

\caption{ 
 $\hat{\rho}_\rmi{test}/\tilde\omega \equiv
 {\rho}_\rmi{test}/(\tilde\omega T^3)$ as a function 
        of $\tilde\omega = \omega/(2\pi T)$,  
        for various relative accuracies $\sigma$, as well as  
        $\ell_\rmii{max}$ chosen from  
        ``windows of opportunity'' in which
        $\sum_{\ell = 0}^{\ell_\rmii{max}} \hat a_\ell$ 
        approximately vanishes, 
        cf.\ \fig\ref{fig:1}
        (results are shown for one random configuration).
        The thick solid line 
        is the correct (input) result.
        The case $\sigma = 10^{-3}$ with $N=48$
        shows a ``failed'' example:
        there is a minimum in 
        $\sum_{\ell = 0}^{\ell_\rmii{max}} \hat a_\ell$ 
        but it is not near zero (cf.\ \fig\ref{fig:1}).
 }
\la{fig:2}
\end{figure*}

The Euclidean correlator is subsequently
integrated numerically from 
\be
 {\mathcal G}(\tau,\cdot) = 
 \int_0^\infty
 \frac{{\rm d}\omega}{\pi} \rho_\rmi{test}
 \frac{\cosh \left(\frac{\beta}{2} - \tau\right)\omega}
 {\sinh\frac{\beta \omega}{2}} 
 \;, \la{master}
\ee
for $\tau = k \beta/N \le \beta/2$. At each $\tau$ 
we add a random error from a Gaussian distribution of relative
variance $\sigma^2$, and then mirror ${\mathcal G}(\tau,\cdot)$ 
to the whole interval through the symmetry in $\tau\to \beta-\tau$.
On this ``data'' the steps of section~4 are applied. 
Small variants of \eq\nr{model} bring along little change, 
but the situation deteriorates rapidly if the structure 
is more peaked either in the ultraviolet 
($\omega \gg 2 \pi T$) 
or in the infrared ($\omega \ll 2 \pi T$).

The behaviour of the coefficients $a_{\ell}$, in particular
the ``diagnostic'' sum of \eq\nr{sumal1}, is illustrated
in \fig\ref{fig:1}; the quality of recovering 
the spectral function using windows of opportunity deduced from
\fig\ref{fig:1} is shown
in \fig\ref{fig:2}. Fig.~\ref{fig:3} demonstrates the
effect of statistical noise on the final results.
The lessons we draw are the following: 
\begin{itemize}

\item[(i)]
The recovery is reasonable only when 
$\sum_{\ell=0}^{\ell_\rmii{max}} a_\ell \approx 0$.
For good accuracy, the sum 
$\sum_{\ell=0}^{\ell_\rmii{max}} a_\ell$ shows a near-zero minimum; 
then $\ell_\rmi{max}$ should be chosen close to the minimum
(the function $\rho(\omega,\cdot)/\omega$ is also extremal there). 
For poor accuracy it could happen that no clear minimum is seen, 
cf.\ $\sigma = 10^{-3}$ in \fig\ref{fig:1}(left); then 
$\ell_\rmi{max}$ should be chosen close 
to a point where $\sum_{\ell=0}^{\ell_\rmii{max}} a_\ell$ 
crosses zero. 

\item[(ii)]
Other values of $\ell_\rmi{max}$ lead in general to nonsensical results. 

\item[(iii)]
Within a given ``robust'' near-zero minimum, the
dependence on $N$ and $\sigma$ is quite mild. 

\item[(iv)]
The recovery can be qualitatively improved only
by increasing the accuracy so much that a second window 
opens up (cf.\ $\sigma = 10^{-8}$ in \fig\ref{fig:1}). This is 
unlikely to be reached in practice and, in any case, the improvement
is not that overwhelming (cf.\ \fig\ref{fig:2}). 

\item[(v)]
The statements above apply to any single random 
configuration. With a sample of them, $\ell_\rmi{max}$ 
could be separately fixed for each configuration. 
Within our toy model, 
it requires an accuracy $\sigma \lsim 10^{-4}$ to find 
a useful minimum for almost every configuration (cf.\ \fig\ref{fig:3}). 
For $\sigma = 10^{-3}$, typical configurations show no near-zero 
minimum (cf.\ \fig\ref{fig:1}), but rare ones do and if
it is possible to restrict the statistics to those and to carry out
the averaging on the final $\rho(\omega,\cdot)/\omega$, 
then a rough estimate can still be obtained.

\end{itemize}

Although it goes beyond the scope of the present note to carry 
out a detailed investigation of issues related to statistical 
analysis, we note that, in general, the output function depends 
non-linearly on input data, because the value of $\ell_\rmii{max}$ 
varies and affects significantly the result. Error estimation 
should therefore be carried out with e.g.\ jackknife or 
bootstrap methods, perhaps with blocked configurations (the effect 
of blocking has been shown to be beneficial in connection 
with the Maximum Entropy Method, see e.g.\ ref.~\cite{condmat1}).

\begin{figure*}


\centerline{%
\epsfysize=7.5cm\epsfbox{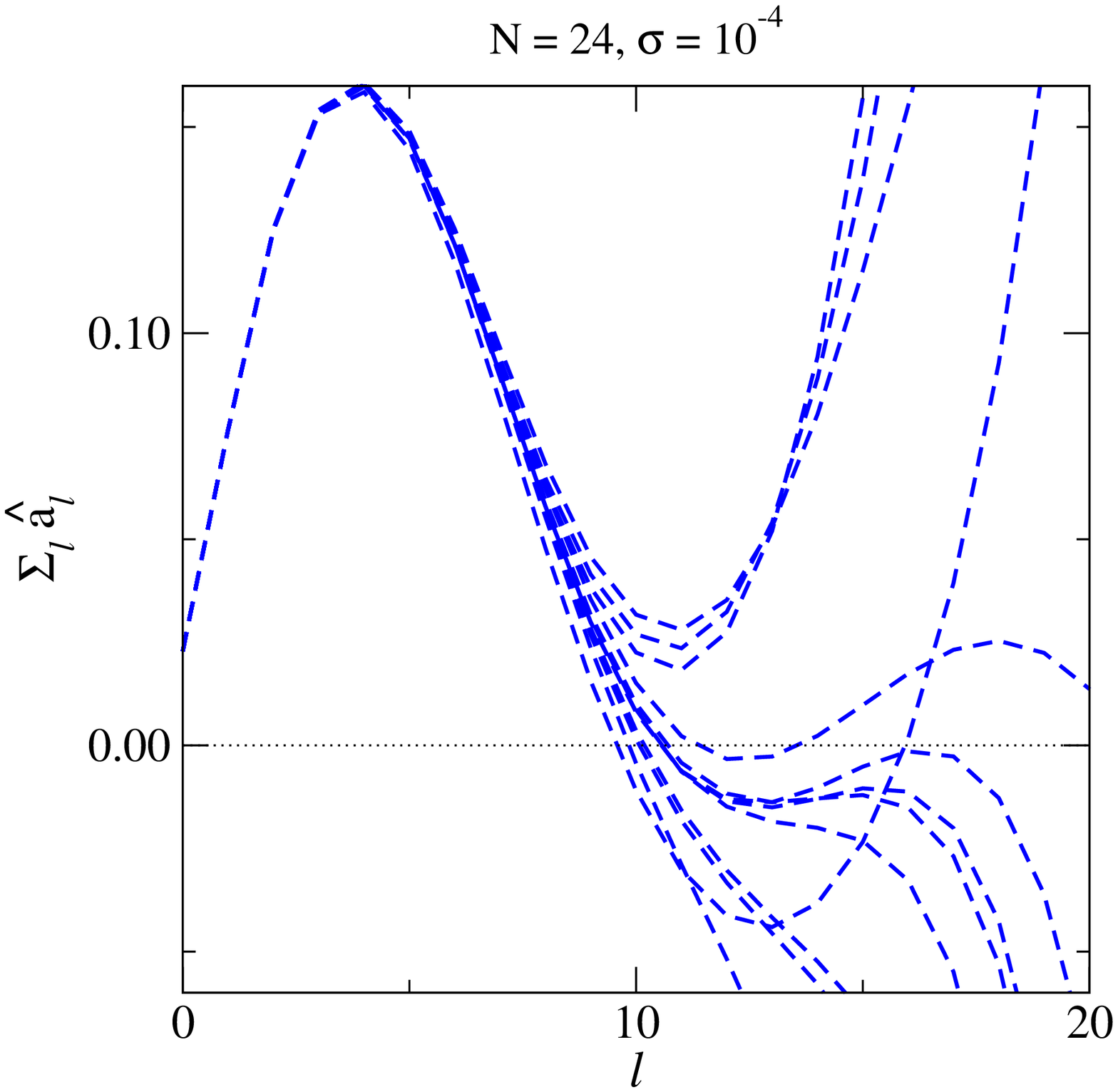}%
~~\epsfysize=7.5cm\epsfbox{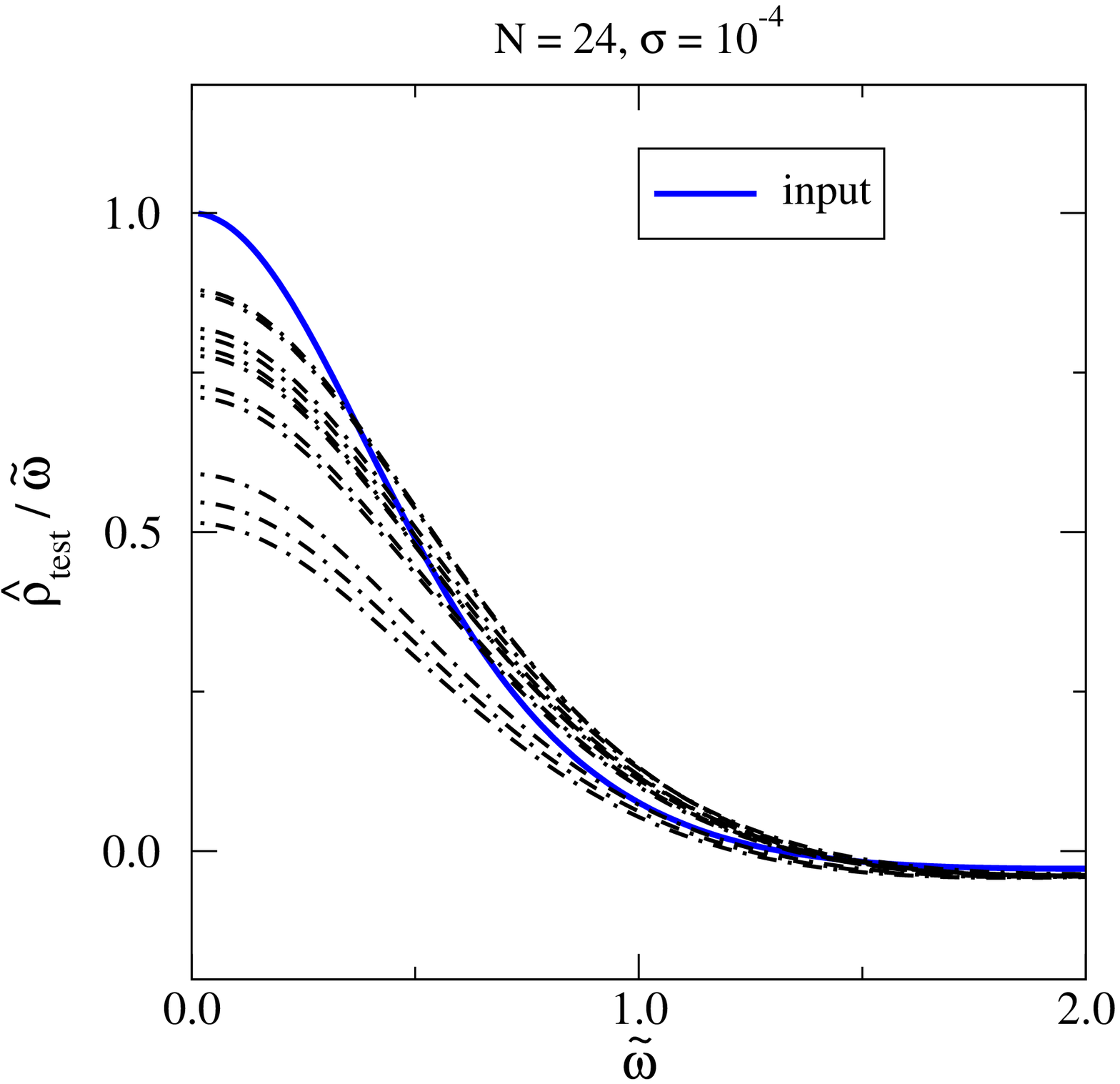}%
}

\caption{ 
 Results from a sample of random configurations 
 at $N=24$, $\sigma = 10^{-4}$, for 
 $\sum_{\ell = 0}^{\ell_\rmii{max}} \hat a_\ell$
 (left)
 and 
 ${\rho}_\rmi{test}/(\tilde\omega T^3)$
 (right).
 The parameter $\ell_\rmii{max}$ was chosen
 according to the criteria specified in the text,
 and varies within the range $10 - 13$. 
 The three lowest-most curves in the right panel correspond
 to the three cases in the left panel in which 
 $\sum_{\ell = 0}^{\ell_\rmii{max}} \hat a_\ell$ does not cross zero;
 omitting such configurations from the statistics would appear to 
 reduce the systematic error of the final average.  
 }
\la{fig:3}
\end{figure*}


%
\section{Conclusions}

%
The algorithm of ref.~\cite{cuniberti} 
possesses a number of attractive features: 
it can be fully specified in a small number of explicit steps; 
it requires no priors; it does not necessitate a positive-definite 
spectral function; and it projects out the Matsubara 
zero-mode contribution whose handling 
has been considered a problem in certain contexts. 

Unfortunately, from a practical point of view,
the algorithm of ref.~\cite{cuniberti} cannot be guaranteed 
to yield a quantitatively accurate analytic continuation 
of thermal imaginary-time data.  
In some sense, the situation is akin to 
the sign problem hampering simulations of QCD with a finite 
baryon number density: there are significant cancellations taking 
place, particularly if a spectral function at a small frequency
$\omega \ll 2 \pi T$ needs to be determined. 
Also, short-distance divergences need to be subtracted from
the Euclidean  correlator $\mathcal{G}(\tau,\cdot)$, 
which constitutes a significance loss of its own. 

Nevertheless, we have demonstrated that in a lucky case with 
a structureless spectral function and precise data 
(with relative errors $< 0.1$\% after the ultraviolet subtraction), 
already $N \gsim 20$ data points may yield a qualitative reproduction
of a transport coefficient (zero-frequency intercept of 
$\rho(\omega,\cdot)/\omega$). 
In general, it is difficult 
to estimate systematic errors, but {\em if} a clear near-zero 
minimum in $\sum_{\ell = 0}^{\ell_\rmii{max}} a_\ell$ is found
as a function of $\ell_\rmi{max}$, then it appears that
a $\lsim 50$\% uncertainty can be expected. This could already 
be useful, given that current model-independent determinations of 
transport coefficients might contain errors of more than 100\%~\cite{chm}.

%
\section*{Acknowledgements}

M.L.\ thanks Dietrich B\"odeker for informing him about
ref.~\cite{cuniberti} several years ago, and Harvey Meyer
for useful discussions.  
M.L.\ was partly supported by 
the BMBF under project
{\em Heavy Quarks as a Bridge between
     Heavy Ion Collisions and QCD}; 
L.M.\ was supported by the Alexander von Humboldt foundation and the Academy of Finland. 


\appendix
\renewcommand{\thesection}{Appendix~\Alph{section}}
\renewcommand{\thesubsection}{\Alph{section}.\arabic{subsection}}
\renewcommand{\theequation}{\Alph{section}.\arabic{equation}}


\end{document}